\RequirePackage{fixltx2e}
\documentclass[10pt, a4paper, conference]{IEEEtran}
\usepackage[T1]{fontenc}
\usepackage{amsmath}
\usepackage{amsthm}
\usepackage{amssymb}
\usepackage{graphicx}
\usepackage[footnotesize, center]{caption2}
\usepackage{type1cm}

\makeatletter
\theoremstyle{plain}

\theoremstyle{plain}

\theoremstyle{remark}

\usepackage[caption=false,font=footnotesize]{subfig}

\makeatother
\providecommand{\lemmaname}{Lemma}
\providecommand{\remarkname}{Remark}
\providecommand{\theoremname}{Theorem}
\IEEEoverridecommandlockouts
\begin{document}

\title{End to End Performance Analysis of Relay Cooperative Communication Based on Parked Cars}

\author{\IEEEauthorblockN{Yingying SUN\IEEEauthorrefmark{1}, Lijun WANG\IEEEauthorrefmark{2}\IEEEauthorrefmark{3},
Zhiquan BAI\IEEEauthorrefmark{4}, Kyung Sup KWAK\IEEEauthorrefmark{5},
Xuming YAO\IEEEauthorrefmark{1} and Tao HAN\IEEEauthorrefmark{1}} 
\IEEEauthorblockA{\IEEEauthorrefmark{1}School of Electronic Information and Communications,
Huazhong University of Science and Technology, China} 
\IEEEauthorblockA{\IEEEauthorrefmark{2}School of Electronic Information, Wuhan University,
China}  
\IEEEauthorblockA{\IEEEauthorrefmark{3}Department of Information Science and Technology,
Wenhua College, China} 
\IEEEauthorblockA{\IEEEauthorrefmark{4}School of Information Science and Engineering,
Shandong University, China} 
\IEEEauthorblockA{\IEEEauthorrefmark{5}Inha Hanlim Fellow Professor, Department of
Information and Communication, Inha university, Korea} 
\textbf{\small{}sunny\_hbqq@hust.edu.cn, wanglj22@163.com, zqbai@sdu.edu.cn,
kskwak@inha.ac.kr, yaoxuming@hust.edu.cn, }\\
\textbf{\small{}hantao@hust.edu.cn }\thanks{The corresponding author is Tao Han. The authors would like 
to acknowledge the support of the International Science and Technology Cooperation Program of China (Grant No. 2015DFG12580), 
the National Natural Science Foundation of China (Grant Nos. 61471180 and 61771291), the Fundamental Research Funds of 
Shandong University (Grant No. 2016JC010), the National Research Foundation of Korea-Grant funded by the Korean Government 
(Ministry of Science, ICT and Future Planning)-NRF-2014K1A3A1A20034987), and the EU FP7-PEOPLE-IRSES (Grant No. 610524). 
This research is supported in part by the China International Joint Research Center of Green Communications and Networking (No. 2015B01008).}}
\maketitle

\noindent
\begin{abstract}
Parking lots (PLs) are usually full with cars. If these cars are formed
into a self-organizing vehicular network, they can be new kind of
road side units (RSUs) in urban area to provide communication data
forwarding between mobile terminals nearby and a base station. However
cars in PLs can leave at any time, which is neglected in the existing
studies. In this study, we investigate relay cooperative communication
based on parked cars in PLs. Taking the impact of the car's leaving
behavior into consideration, we derive the expressions of outage probability
in a two-hop cooperative communication and its link capacity. Finally,
the numerical results show that the impact of a car's arriving time
is greater than the impact of its parking duration on outage
probability.\\
\end{abstract}

\noindent

\section{Introduction}

Vehicular ad hoc network (VANET) is an important part of the future smart
city, which can support services such as data forwarding, content
caching and mobile edge computation \cite{han_5g_2017,wang_cell-less_2017}. Studies on vehicular network
in the past mainly concentrated on moving cars. With the number of worldwide vehicles
increasing rapidly, and most of cars spend most of the time in
a day on parking, there are large numbers of parked cars in urban areas. Regarding
these parked vehicles as an infrastructure, we can make use of their
own wireless devices and CPUs to improve the communication quality
and computation ability in urban vehicular network \cite{hou_vehicular_2016}.
Some studies have proposed the idea of introducing parked cars into
vehicular networks as static nodes for relay forwarding \cite{liu_pva_2011,eckhoff_cooperative_2011}
and content caching \cite{malandrino_role_2014,zhao_vadd:_2006,liu_sharing_2012}.
In the intelligent transportation system, moving vehicles on the road
always need to share safety message with each other. Studies show that
there will take at least 100 seconds to broadcast important safety
message to all nearby cars, if in areas of low car density \cite{viriyasitavat_uv-cast:_2011}.
One way to overcome it is to deploy roadside units (RSUs) alongside
roads, which brings high installation and maintenance costs. Therefore,
some researchers proposed the idea of regarding parked cars as temporary
RSUs \cite{reis_parked_2017,tonguz_cars_2013}. Besides, parked cars
can also be used as computation nodes in urban computing \cite{hou_vehicular_2016}.
Parked cars are usually regarded as natural roadside nodes, because
they are large number, long-time staying, wide distribution, and specific
location \cite{liu_pva_2011}. However, there are another two features
of parked cars that they are energy-constrained and can leave at any
time, which have been neglected by most studies before.

When a car serves as a relay node, it may leave at any time during a period of communication.
Once a car leaves, the communication link in which it serves as a
relay node becomes disconnected immediately. For point-to-point communication,
the use of multi-relay cooperation can reduce the impact of car's
departure behavior on communication quality. When a car serves as
a computing node, its departure will make computing sub-task interrupted,
leading to the failure of the whole computing task. Therefore, the
impact of car's leaving behavior cannot be ignored.

In this study, we focus on the impact of cars' departure behavior.
We choose cars in a parking lot as study objects. Because parking lots
are generally located in the business district or residential area,
where the communication needs are higher than remote areas. In addition, the vehicle density
in parking lots is maintained at a high level for a long time. We
can form clusters with vehicles close to each other, which can provide
services for moving cars and pedestrians nearby. We combine the probability
distribution of parking duration in \cite{reis_statistics_2016} with
the probability distribution of car's arriving time in \cite{guner_impact_2016}
to describe car's leaving behavior, whose impacts are introduced into
the derivation of outage probability and channel capacity.

The rest of this paper is organized as follows: Section II reviews
related work. Section III describes the system model. The derivations
of outage probability expression and link capacity expression are
presented in Section IV. Section V shows the numerical results and
discusses the impacts of leaving behavior on outage probability. Finally
Section VI concludes this study and proposes future work.

\section{Related Work}

At present, studies of parked cars mainly focus on relay forwarding
and content caching. The idea of introducing parked cars into vehicular
network has been proposed in \cite{liu_pva_2011}, the authors found
that even if a fraction of assistant parked cars could improve network
connectivity significantly. The work in \cite{eckhoff_cooperative_2011}
made use of parked cars at key interaction points to overcome the
signal degradation caused by buildings in the line of sight of two
cars. Eckhoff et al. \cite{eckhoff_cooperative_2011} introduced a
novel scheme to emit cooperative awareness messages at low vehicle
densities periodically, and used the Manhattan grid model to simulate
the street environment as well as gave the obstacle model. In \cite{reis_parked_2017},
Reis et al. found parked cars could indeed serve both as RSUs and
an extension to vehicular infrastructure deployments. Besides, the
work investigated how to select a suitable parked car to be an RSU,
and proposed an on-line algorithm to maximize the coverage by activating
the least parked cars. Su et al. \cite{su_novel_2014} proposed a novel
algorithm which utilized parked cars to cache content based on access
pattern of content. The authors of \cite{su_game_2016} combined RSUs
with parked cars to cache content, used Stackelberg game model to
decide what content would be obtained from RSUs or parked cars, which
minimized the expense of content delivery.

Studies above all didn't take the impact of car's departure into account.
In \cite{reis_statistics_2016}, Reis et al. provided statistic analysis
about survey data from the three metropolitan areas and extracted
the parking behavior of individuals vehicles. The author modeled the
time that a car would spend on parking as a dual Gamma stochastic process
and gave the probability distribution function of it. According to
real car arrival/departure data of a representative parking lot (PL),
Guner et al. in \cite{guner_impact_2016}, simulated the car's arriving
times by using Weibull distribution function. In this study, we exploit
the models proposed in \cite{reis_statistics_2016,guner_impact_2016}
to obtain car's departure pattern, which will be applied to the overall
system to analyze the impact of departure behavior on system performance. 

\section{System Model}

This study focuses on communications between mobile terminals nearby
a parking lot and a base station. The mobile terminals include pedestrians
and slow-moving vehicles because of heavy traffic, which always have
needs of long-time communication. Assuming every car in PLs is equipped
with one antenna, those cars form clusters according to the distribution
of their physical location. Each cluster has one cluster head (CH),
which stores arriving time and the time its cluster members have parked.
Parked vehicles themselves have no demand for communication, only forward
data. Therefore, parked cars in a cluster only send and receive a
small amount of control information among each other, for keeping
the channels synchronized and transmitting channel state information
(CSI). We assume that all cooperative parked cars forward identical
data to obtain space diversity gain.
\begin{figure}[tbh]
\begin{centering}
\includegraphics[width=9cm]{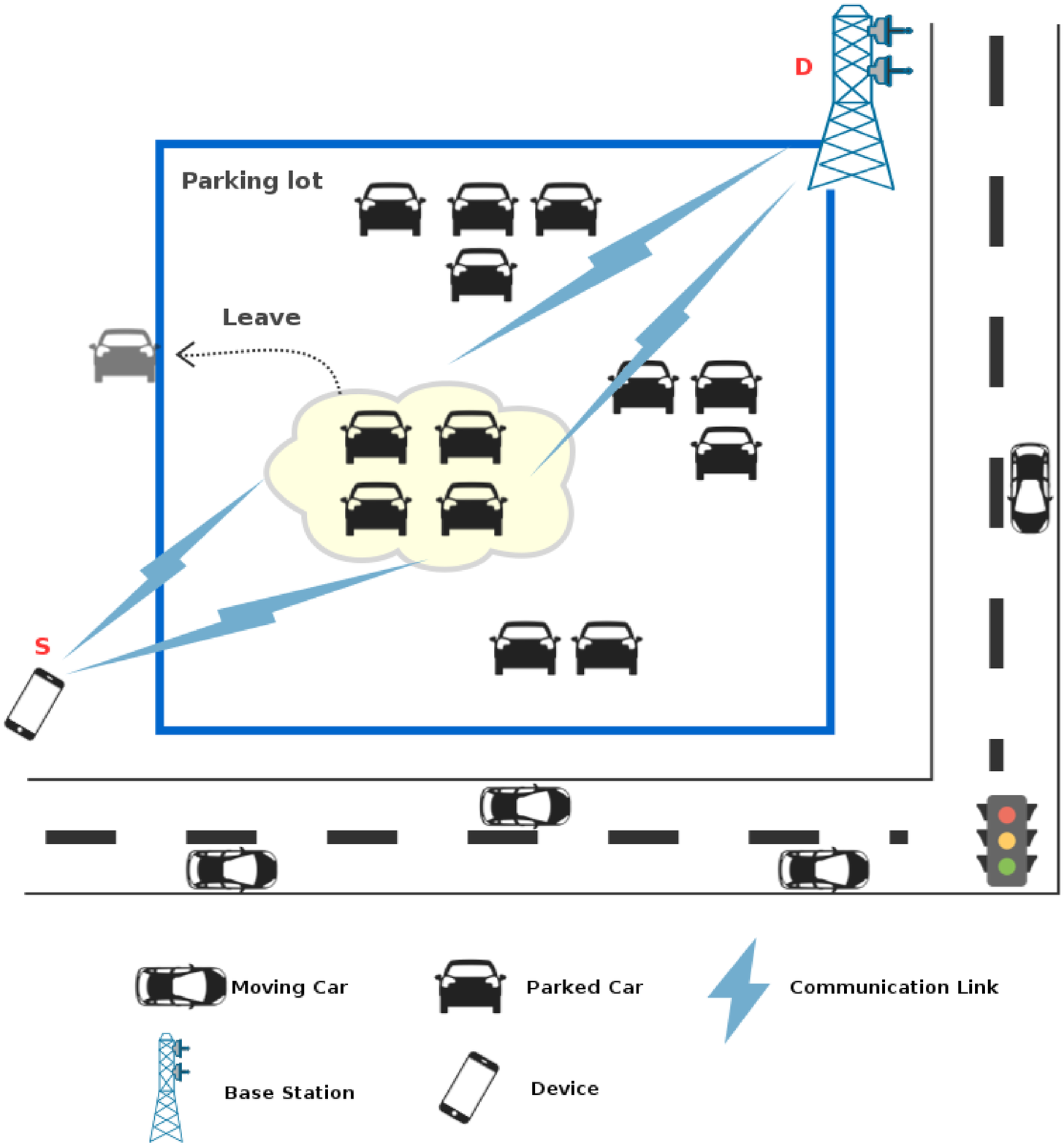}
\par\end{centering}
\caption{\label{fig:1} System Model}

\end{figure}

Consider the cooperative relays system illustrated as Figure \ref{fig:1}.
Assuming a device needs to communicate with the nearby base station, the
device is the transmitter, namely $S$ in Figure \ref{fig:1}, base
station $D$ is the receiver. There is no direct link between $S$
and $D$. The transmitter selects one cluster according to information
stored by CHs and chooses $K(K\geq1)$ parked cars
from the selected cluster. We consider half-duplex mode of operation.
Relay forwarding has two phases. In phase one, the transmitter $S$
broadcasts the signal $x$ with the transmitting power $P_{\mathrm{s}}$ , the receive signal at $i$-th relay is given
by
\begin{equation}
y_{\mathrm{sri}}=\sqrt{P_{\mathrm{s}}}h_{\mathrm{sri}}x+n_{\mathrm{sri}}\:\:i\in\left[1,2,...,K\right].
\end{equation}
where $h_{\mathrm{sri}}$ is the channel coefficient from $S$ to the $i$-th relay and $n_{\mathrm{sri}}$ denotes
the additive white Gaussian noise (AWGN) at the $i$-th relay.

In phase two, $K$ relays amplify the identical signal and forward
it to the receiver $D$, the receiving signal at $D$ from the $i$-th
relay is given by
\begin{equation}
y_{\mathrm{rid}}=h_{\mathrm{rid}}\beta_{i}y_{\mathrm{sri}}+n_{\mathrm{rid}}\:\:i\in\left[1,2,...,K\right],
\end{equation}
where $h_{\mathrm{rid}}$ is the channel coefficient from the $i$-th relay to $D$ and $n_{\mathrm{rid}}$ denotes
the AWGN at $D$.
The relay channels are represented by $h_{\mathrm{sri}}$ and $h_{\mathrm{rid}}$,
which are circularly symmetric complex Gaussian random variables with
unit variance and zero mean. And $n_{\mathrm{sri}},n_{\mathrm{rid}}\sim\mathcal{C}\mathcal{N}\left(0,N_{0}\right)$.
$\beta_{i}$ is an amplification factor that ensures the transmitting
power at $i$-th relay falls below a specific value $P_{\mathrm{ri}}$,
which is given by
\begin{equation}
\beta_{i}=\sqrt{\frac{P_{\mathrm{ri}}}{P_{\mathrm{s}}|h_{\mathrm{sri}}|^{2}+N_{0}}}.
\end{equation}

For single relay link, the signal-to-noise ratio (SNR) at the $i$-th
relay can be represented by\cite{emamian_multi-user_2002}
\begin{equation}
\gamma_{i}=f\left(\frac{P_{\mathrm{s}}|h_{\mathrm{sri}}|^{2}}{N_{0}},\frac{P_{\mathrm{ri}}|h_{\mathrm{rid}}|^{2}}{N_{0}}\right),\label{eq:4}
\end{equation}
where $f\left(x,y\right)=\frac{xy}{1+x+y}$, and $|h_{\mathrm{sri}}|^{2},|h_{\mathrm{rid}}|^{2}\sim E\left(\frac{1}{2}\right)$.

The arrival time $t$ of vehicles can be described by a Weibull distribution \cite{guner_impact_2016},
which is given by
\begin{equation}
f\left(t\right)=\frac{\alpha}{\beta}\left(\frac{t}{\beta}\right)^{\alpha-1}e^{-\left(\frac{t}{\beta}\right)^{\alpha}},\label{eq:5}
\end{equation}
where $\alpha$ and $\beta$ denote shape and scale parameters. In \cite{guner_impact_2016},
$\alpha=0.9831$ and $\beta=16.8$. The time $x$ a car will spend
on parking follows a dual Gamma distribution \cite{reis_statistics_2016},
when it arrives at hour $t$. The probability density function (PDF)
is given by
\begin{align}
f\left(x,t\right) & =\frac{D_{1,t}}{\Gamma\left(\kappa_{s,t}\right)\theta_{s,t}^{\kappa_{s,t}}}x^{\kappa_{s,t}-1}e^{-\frac{x}{\theta_{s,t}}}\nonumber \\
 & +\frac{D_{2,t}}{\Gamma\left(\kappa_{l,t}\right)\theta_{l,t}^{\kappa_{l,t}}}x^{\kappa_{l,t}-1}e^{-\frac{x}{\theta_{l,t}}},\nonumber \\
 & \qquad x>0,\:t=\left\{ 0,1,2,...,23\right\} ,\label{eq:6}
\end{align}
where $\left\{ \kappa_{s,t},\theta_{s,t},\kappa_{l,t},\theta_{l,t},D_{1,t},D_{2,t}\right\} $
are specific to car's arrival time $t$, and their values can be
seen in the appendix of \cite{reis_statistics_2016}. $\Gamma\left(\cdot\right)$
is the Gamma function.

Different from traditional static relays, parked cars may leave at any time. Therefore, for a single relay link, there are two cases where an interrupt will occur:
(1) the SNR of the $i$-th link is below the threshold, whose probability is expressed
as $P_{i}^{\mathrm{\mathrm{\gamma}}}$; (2) the $i$-th relay leaves
in the communication process, whose probability is expressed as $P_{i}^{\mathrm{L}}$.

\section{Analysis of Outage Probability and Channel Capacity}

In this section, we analyze the impact of leaving behavior on outage
and channel capacity. First, we give the expression of $P_{i}^{\mathrm{L}}$. Then, we derive the analytical expressions
of outage and channel capacity on the basis of the above analysis.

\subsection{Outage Probability of The System}

In a period of communication time, relay's departure will cause the
link to be interrupted. According to \eqref{eq:6}, the probability that
a vehicle still has at least $n$ hours left for parking\cite{reis_statistics_2016},
when it has been parked for $t_{\mathrm{a}}$ time, is derived as
\begin{figure*}[tbh]
\noindent 
\begin{equation}
P\left[X_{\mathrm{t}}>t_{\mathrm{a}}+n|X_{\mathrm{t}}>t_{\mathrm{a}}\right]=\frac{D_{1,t}\gamma\left(\kappa_{s,t},\frac{t_{\mathrm{a}}+n}{\theta_{s,t}}\right)\Gamma\left(\kappa_{l,t}\right)+D_{2,t}\gamma\left(\kappa_{l,t},\frac{t_{\mathrm{a}}+n}{\theta_{l,t}}\right)\Gamma\left(\kappa_{s,t}\right)-\Gamma\left(\kappa_{l,t}\right)\Gamma\left(\kappa_{s,t}\right)}{D_{1,t}\gamma\left(\kappa_{s,t},\frac{t_{\mathrm{a}}}{\theta_{s,t}}\right)\Gamma\left(\kappa_{l,t}\right)+D_{2,t}\gamma\left(\kappa_{l,t},\frac{t_{\mathrm{a}}}{\theta_{l,t}}\right)\Gamma\left(\kappa_{s,t}\right)-\Gamma\left(\kappa_{l,t}\right)\Gamma\left(\kappa_{s,t}\right)}\label{eq:7}
\end{equation}

\rule[0.5ex]{1\textwidth}{0.5bp}
\end{figure*}
 \eqref{eq:7}, where $\Gamma\left(\cdot\right)$ and $\gamma\left(\cdot\right)$
are the upper and lower incomplete Gamma functions respectively. Then
the probability of a car has at most $n$ hours left can be expressed
by
\begin{equation}
P\left[X_{\mathrm{t}}\leq t_{\mathrm{a}}+n|X_{\mathrm{t}}>t_{\mathrm{a}}\right]=1-P\left[X_{\mathrm{t}}>t_{\mathrm{a}}+n|X_{\mathrm{t}}>t_{\mathrm{a}}\right].\label{eq:8}
\end{equation}

It can be seen in \eqref{eq:8} that $n=0$ indicates the car leaves
at the moment, the probability of which is always zero. But cars are
possible to leave at any time.We try two ways to solve this contradiction:
\begin{enumerate}
\item Use the probability of leaving within a short time $\tau$ to represent
the leaving probability at the moment. Then the probability that relay $i$
leaves at the moment can be expressed by
\begin{equation}
P_{i}^{\mathrm{L}}=1-P\left[X_{\mathrm{t}}>t_{\mathrm{dur}}+\tau|X_{\mathrm{t}}>t_{\mathrm{dur}}\right].\label{eq:9}
\end{equation}
\item Now that we have the PDFs of the time that a car arrives at a PL and
a car's parked duration, we can set each car's arrival time and parked
duration according to \eqref{eq:5}, \eqref{eq:6}. Then we can use Monte
Carlo simulation to analyze the impact of departure behavior on outage
probability.
\end{enumerate}

For a single relay link, the cumulative distribution function (CDF)
of its SNR is given in \cite{hasna_end--end_2003}
\begin{align}
F_{\gamma_{i}}\left(x|w_{\mathrm{sri}},w_{\mathrm{rid}}\right) & =1-2x\sqrt{w_{\mathrm{sri}}w_{\mathrm{rid}}}e^{-x\left(w_{\mathrm{sri}}+w_{\mathrm{rid}}\right)}\nonumber \\
 & \times K_{1}\left(2x\sqrt{w_{\mathrm{sri}}w_{\mathrm{rid}}}\right),\label{eq:10}
\end{align}
where $K_{1}\left(\cdot\right)$ is the first order modified Bessel
function of the second kind, $w_{\mathrm{sri}}$ and $w_{\mathrm{rid}}$ are the reciprocals of the average SNRs of the two links respectively, one is from $S$ to the $i$-th relay 
and the other is from the $i$-th relay to $D$. Then $w_{\mathrm{\mathrm{sri}}}=\frac{N_{0}}{2P_{\mathrm{s}}}$,
$w_{\mathrm{rid}}=\frac{N_{0}}{2P_{\mathrm{ri}}}$. Because the $K_{1}\left(x\right)$
function converges to $1/x$ when $x$ approaches zero, the formula
\eqref{eq:10} can be simplified to
\begin{equation}
F_{\gamma_{i}}\left(x|\mu_{i}\right)=1-e^{-\mu_{i}x},
\end{equation}
where $\mu_{i}=w_{\mathrm{sri}}+w_{\mathrm{rid}}$. The probability
that the SNR of the $i$-th link is below the threshold $\gamma_{\mathrm{th}}$
can be expressed as
\begin{equation}
P_{i}^{\gamma}=F_{\gamma_{i}}\left(\gamma_{\mathrm{th}}|\mu_{i}\right).
\end{equation}

On the basis of above analysis, the outage probability of the $i$-th
relay link is given by
\begin{align}
P_{i}^{\mathrm{out}} & =f\left(P_{i}^{\gamma},P_{i}^{\mathrm{L}}\right)\nonumber \\
 & =1-(1-P_{i}^{\gamma})\left(1-P_{i}^{\mathrm{L}}\right).\label{eq:13}
\end{align}
To simplify the analysis, we choose select combination (SC) at the receiver. Thus,
the outage probability of each link is independent of each other.
And the outage probability of the two-hop communication link can be
expressed as
\begin{equation}
P_{\mathrm{out}}=\underset{i\in\varOmega}{\prod}P_{i}^{\mathrm{out}}.
\end{equation}
where $\varOmega$ is the cooperative relay set.

\subsection{Capacity of The System}

We use select combination technology at the receiver for simplicity,
and the SNR at the receiver is given by
\begin{equation}
\gamma_{\mathrm{sc}}=\underset{i\in\varOmega}{max}\gamma_{i},
\end{equation}
where $\varOmega$ is the cooperative relay set and $\gamma_{i}$ is the SNR of the $i$-th relay link. When the number of
relays is $K$, the CDF of $\gamma_{\mathrm{sc}}$ can be expressed as
\begin{align}
F_{\gamma_{\mathrm{sc}}}\left(\gamma\right) & =P_{r}\left\{ \gamma_{1}<\gamma,\gamma_{2}<\gamma,...,\gamma_{\mathrm{K}}<\gamma\right\} \nonumber \\
 & =\overset{{\scriptstyle K}}{\underset{{\scriptstyle i=1}}{\prod}}F_{\gamma_{i}}\left(\gamma|w_{\mathrm{sri}},w_{\mathrm{rid}}\right).\label{eq:16}
\end{align}

Then the PDF can be obtained by deriving from the formula \eqref{eq:16}.
Put \eqref{eq:8} into \eqref{eq:16} and we have

\begin{align}
f_{\gamma_{\mathrm{sc}}}\left(\gamma\right) & =\overset{{\scriptstyle K}}{\underset{{\scriptstyle i=1}}{\sum}}\left\{ f_{\gamma_{i}}\left(\gamma\right)\overset{{\scriptstyle K}}{\underset{{\scriptstyle j=1,j\neq i}}{\prod}}F_{\gamma_{i}}\left(\gamma|w_{\mathrm{sri}},w_{\mathrm{rid}}\right)\right\} \nonumber \\
 & =\overset{{\scriptstyle K}}{\underset{{\scriptstyle i=1}}{\sum}}\left\{ \mu_{i}e^{-\mu_{i}\gamma}\overset{{\scriptstyle K}}{\underset{{\scriptstyle j=1,j\neq i}}{\prod}}\left(1-e^{-\mu_{i}\gamma}\right)\right\} .\label{eq:17}
\end{align}
according to \eqref{eq:8}, $\mu_{i}=w_{\mathrm{sri}}+w_{\mathrm{rid}}=\frac{N_{0}}{2P_{\mathrm{s}}}+\frac{N_{0}}{2P_{\mathrm{ri}}}$.
We set $P_{\mathrm{ri}}=P_{\mathrm{s}}$ for all relays, then $\mu_{i}=\bar{\mu}=\frac{N_{0}}{P_{\mathrm{s}}}$,
for all $i\in\left\{ 1,2,...,K\right\} $, and we can simplify the
formula \eqref{eq:17} to
\begin{equation}
f_{\gamma_{\mathrm{sc}}}\left(\gamma\right)=K\bar{\mu}e^{-\bar{\mu}\gamma}\left(1-e^{-\bar{\mu}\gamma}\right)^{K-1}.
\end{equation}

When the transmitting power is constant, the channel capacity with
SC can be expressed by
\begin{align}
C & =\frac{1}{2}B\int_{\gamma}^{\infty}\mathrm{log_{2}}\left(1+\gamma\right)f_{\gamma_{\mathrm{sc}}}\left(\gamma\right)d\gamma\nonumber \\
 & =\frac{BK}{2\mathrm{ln}2}\overset{{\scriptstyle K}}{\underset{{\scriptstyle i=1}}{\sum}}\left(-1\right)^{i}\left(\begin{array}{c}
K-1\\
i
\end{array}\right)e^{\left(i+1\right)\bar{\mu}}\frac{E_{1}\left(\left(i+1\right)\bar{\mu}\right)}{i+1},\label{eq:19}
\end{align}
according to \cite{alouini_capacity_1999}. The coefficient 1/2 is
caused by two time slots occupied by one transmission, and $E_{1}\left(x\right)=\int_{1}^{\infty}t^{-1}e^{-tx}dt, x>0$,
$\left(\begin{array}{c}
K-1\\
i
\end{array}\right)=\frac{\left(K-1\right)!}{\left(K-1-i\right)!i!}$.

Due to car's leaving behavior, the number of relays will be changing
during the communication. We assume that new relays are not allowed
to be added during a period of communication. After a short time $\tau$,
the number of relays is $K^{'}$, and we set $p_{i}$ as the probability that the $i$-th relay will not leave within a short time $\tau$, which equals to $1-P_{i}^{\mathrm{L}}$. 
Because those cars' departure events are independent to each other, we can
obtain the probability distribution of $K^{'}$

\begin{eqnarray}
P\left\{ K^{'}=1\right\}  & = & \overset{{\scriptstyle K}}{\underset{{\scriptstyle i=1}}{\sum}}p_{i}\overset{{\scriptstyle K}}{\underset{{\scriptstyle n=1,n\neq i}}{\prod}}\left(1-p_{n}\right)\nonumber \\
P\left\{ K^{'}=2\right\}  & = & \overset{{\scriptstyle K}}{\underset{{\scriptstyle i=1}}{\sum}}\overset{{\scriptstyle K}}{\underset{{\scriptstyle j=2}}{\sum}}p_{i}p_{j}\overset{{\scriptstyle K}}{\underset{{\scriptstyle n=1,n\neq i,j}}{\prod}}\left(1-p_{n}\right)\nonumber \\
 & \vdots\nonumber \\
P\left\{ K^{'}=k\right\}  & = & \overset{{\scriptstyle K}}{\underset{{\scriptstyle i_{1}=1}}{\sum}}\overset{{\scriptstyle K}}{\underset{{\scriptstyle i_{2}=i_{1}+1}}{\sum}}\cdots\overset{{\scriptstyle K}}{\underset{{\scriptstyle i_{k}=i_{k-1}+1}}{\sum}}p_{i_{1}}p_{i_{2}}\cdots p_{i_{k}}\nonumber \\
 & \times & \overset{{\scriptstyle K}}{\underset{{\scriptstyle \begin{array}{c}
n=1\\
n\neq i_{1},i_{2},\cdots,i_{k}
\end{array}}}{\prod}}\left(1-p_{n}\right).\label{eq:20}
\end{eqnarray}

Taking the car's leaving into consideration, the channel capacity
$C_{l}$ after time $\tau$ can be written as
\begin{equation}
C_{l}=\overset{{\scriptstyle K}}{\underset{{\scriptstyle k=1}}{\sum}}P\left\{ K^{'}=k\right\} C.\label{eq:21}
\end{equation}
\begin{figure*}[t]
\noindent 
\begin{equation}
C_{l}=\overset{{\scriptstyle K}}{\underset{{\scriptstyle k=1}}{\sum}}\left[\left(\overset{{\scriptstyle K}}{\underset{{\scriptstyle i_{1}=1}}{\sum}}\overset{{\scriptstyle K}}{\underset{{\scriptstyle i_{2}=i_{1}+1}}{\sum}}\cdots\overset{{\scriptstyle K}}{\underset{{\scriptstyle i_{k}=i_{k-1}+1}}{\sum}}p_{i_{1}}p_{i_{2}}\cdots p_{i_{k}}\right)\overset{{\scriptstyle K}}{\underset{{\scriptstyle \begin{array}{c}
n=1,\\
n\neq i_{1},\cdots,i_{k}
\end{array}}}{\prod}}\left(1-p_{n}\right)\frac{Bk}{2\mathrm{ln}2}\overset{k}{\underset{{\scriptstyle j=1}}{\sum}}\left(-1\right)^{j}\left(\begin{array}{c}
k-1\\
j
\end{array}\right)e^{\left(j+1\right)\bar{\mu}}\frac{E_{1}\left(\left(j+1\right)\bar{\mu}\right)}{j+1}\right]\label{eq:22}
\end{equation}

\rule[0.5ex]{1\textwidth}{0.5bp}
\end{figure*}

Substitute \eqref{eq:19} and \eqref{eq:20} into \eqref{eq:21},
we obtain the complete expression of the channel capacity as \eqref{eq:22}.

\section{Numerical Result}

This section shows the numerical evaluation of the results presented in
the previous sections, followed by discussions and conclusions. The
typical values of system parameters are selected from certain practical
scenarios. Especially, the time $t_{\mathrm{dur}}$ that a car has
been parked for can be calculated by the subtraction of the arrival
time $t_{\mathrm{arr}}$ and current time $t_{\mathrm{cur}}$, we
set $t_{\mathrm{cur}}=3:00\,\mathrm{P.M}$ and $t_{\mathrm{arr}}$
is the expected value of the arrival time. The transmitting power
of the sender and relays is identical $P_{\mathrm{s}}=P_{\mathrm{ri}}=2\,\mathrm{W}$.

\subsection{Outage Probability}

\begin{figure}
\begin{centering}
\includegraphics[width=9cm]{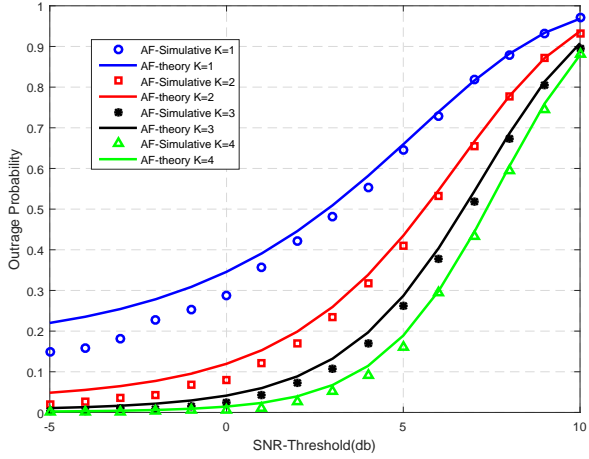}
\par\end{centering}
\caption{\label{fig:2}Outage probability versus threshold SNR for a system}
\end{figure}

Figure \ref{fig:2} shows the outage probability trend versus SNR
threshold. To verify the derivation of the outage probability, we
have also conducted a simulation. The result of the simulation is presented
in Figure \ref{fig:2} with scattering points.
 It is clear from the figure that for all practical purposes the two
curves are essentially indistinguishable and the figure also shows
the improvement in outage probability as the number of the relays
increases. With multi-relays, the impact of car's departure behavior
can be reduced. And the using of multi-relays can reduce the transmit power of
single relay in the same communication quality comparing to the using of single relay. 
Because cars in PLs belong to private owners, the energy
consumption of each car should be as little as possible.

\begin{figure}[tbh]
\begin{centering}
\includegraphics[width=9cm]{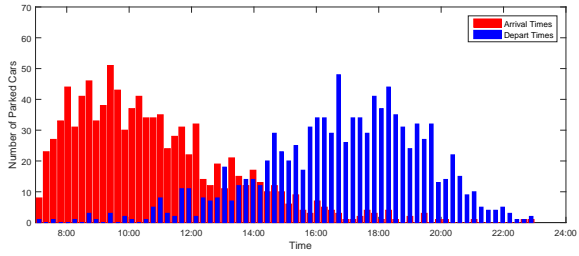}
\par\end{centering}
\caption{\label{fig:3}Arrival/departure times of the cars for a weekday}
\end{figure}

\begin{figure}[tbh]
\begin{centering}
\includegraphics[width=9cm]{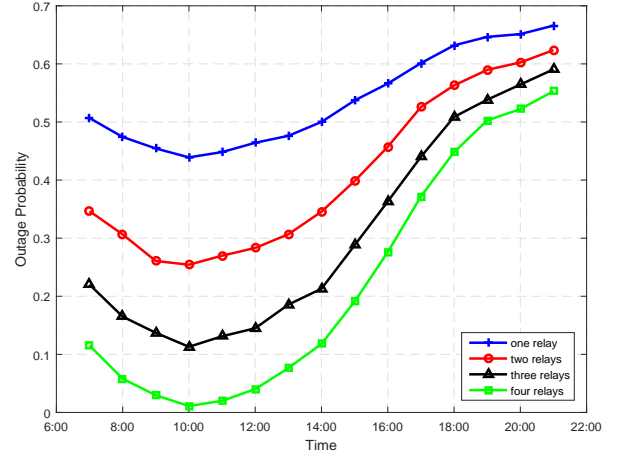}
\par\end{centering}
\caption{\label{fig:4}Outage probability in a day }
\end{figure}

On the basis of PDFs of arriving time in \eqref{eq:5} and the time
a car will spend on parking in \eqref{eq:6} , Figure \ref{fig:3}
presents the arrival/departure times for a sample day. It is seen
that a large number of vehicles arrive at a PL in the morning and
leave at sunset. We guess there may be a certain relationship between
the system's outage probability and daily change of PLs. 
We assume a PL has a total parking capacity of 500 cars, which serves 2,000 cars within a day, and set the arrival time and predicted
parked duration of each car by formulas, \eqref{eq:5} and \eqref{eq:6}.
Then we can obtain the change of outage probability in a day, which
is shown in Figure \ref{fig:4}. As can be seen from the curves, the
outage probability decreases first and then increases. Combining with
Figure \ref{fig:3}, there are lots of cars arriving at the PL but
few cars leaving from 8:00 AM to 11:00 AM. The communication link
established during this period can keep active for a long time, because
these cars have a long-term stay trend. Besides, the number of optional
relays has increased. Thus, the outage probability decreases. Otherwise,
there are lots of cars leaving from the PL but few cars arriving after
5:00 PM. During this period, the outage is easy to happen because
many cars are likely to leave in the short term.

\subsection{Impact of Departure Pattern Parameters}

\begin{figure}[tbh]
\begin{centering}
\includegraphics[width=9cm]{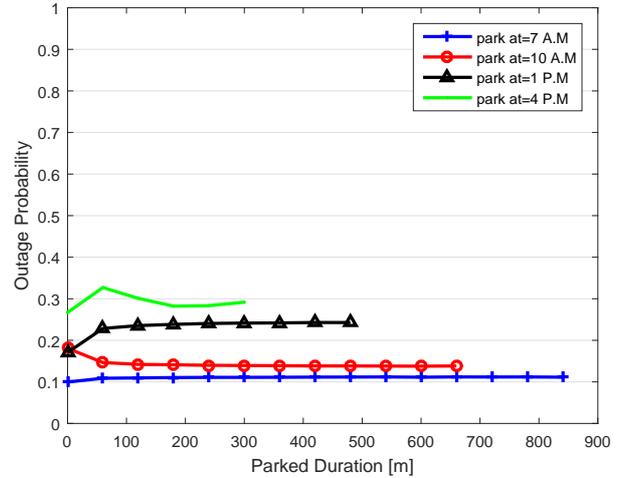}
\par\end{centering}
\caption{\label{fig:5}The outage probability versus $t_{\mathrm{dur}}$, in
minutes}
\end{figure}

In this study, the departure pattern we established has two key parameters,
the arrival time $t_{\mathrm{arr}}$ and the time $t_{\mathrm{dur}}$
a car has parked. The two parameters directly determine the probability
that a car leaves at some time, thus affect the outage probability.
In order to find out the impact of $t_{\mathrm{arr}}$ and $t_{\mathrm{dur}}$,
we calculate the outage probability with only one relay forwarding
data. First, we set a fixed $t_{\mathrm{arr}}$, and calculate $P_{\mathrm{out}}$
with different $t_{\mathrm{dur}}$. Figure \ref{fig:5} shows the
impact of $t_{\mathrm{dur}}$ on the outage probability. It can be
seen that the curves almost parallel to the horizontal axis, which
indicates that $t_{\mathrm{dur}}$ has little impact on $P_{\mathrm{out}}$,
even can be ignored.

\begin{figure}[tbh]
\begin{centering}
\includegraphics[width=9cm]{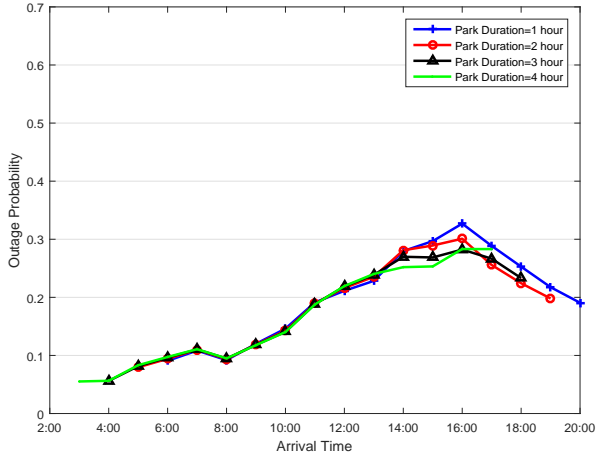}
\par\end{centering}
\caption{\label{fig:6}The outage probability versus $t_{\mathrm{arr}}$}
\end{figure}

Then we set $t_{\mathrm{dur}}$ fixed, the impact of $t_{\mathrm{arr}}$
on the outage probability is shown in Figure \ref{fig:6}. We can
find that selecting cars that arrive at PLs from 4:00 P.M to 6:00
P.M as relays will lead to higher outage probability, because these
cars are likely to leave in short time, while cars that arrive at
earlier time are long-staying in PLs. Therefore, relay selection algorithms
based on car's leaving behavior, which will be presented in the future
work, should mainly focus on the impact of $t_{\mathrm{arr}}$, not
the impact of $t_{\mathrm{dur}}$.

\subsection{Capacity}

\begin{figure}[tbh]
\begin{centering}
\includegraphics[width=9cm]{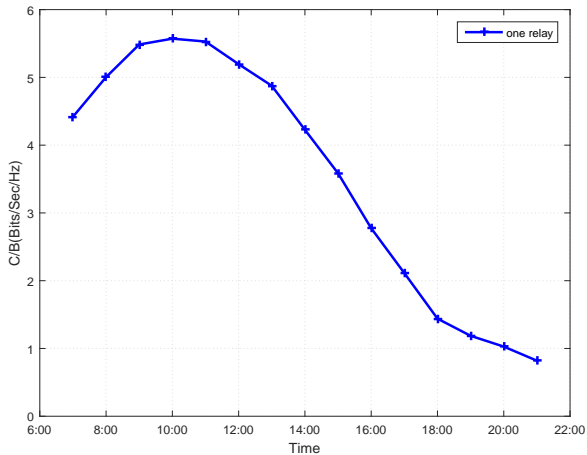}
\par\end{centering}
\caption{\label{fig:7}The channel capacity within a day}

\end{figure}

To find out the change trend of the channel capacity within a day,
we assume there is a PL same as the one mentioned in subsection A,
and select one relay randomly every hour, then we can obtain the change
of the channel capacity in a day. Figure \ref{fig:7} shows the variation
of the channel capacity from 7:00 A.M to 22:00 A.M. We find out that
the channel capacity increases first and then decreases, which is
opposite to the change trend of outage probability.

\section{Conclusions and Future Work}

This study considers a two-hop communication assisted by parked cars
in a PL. We give an appropriate expression of car's leaving probability
based on existing parking patterns. Taking SNR and cars' departure
behavior into account, we derive the expressions of outage probability
and channel capacity of this system. With the analysis of the change
of outage probability within a day, we find that outage probability
of this system is related to the traffic volume in PLs. There are
two peak points of traffic volume at around 8:00 A.M and 19:00 A.M.
From 8:00 A.M to 10:00 A.M, with lots of cars pouring in PLs, the
number of optional cars as relays is increasing. Thus, the outage
probability is low. After 4:00 P.M, with lots of cars leaving from
PLs, the outage probability is increasing. We also investigate the
effects caused by the parameters of the parking model, and find that
the time a car has been parked for has little impact on the outage
probability of this system while the impact of car's arrival time
is greater.

In our system model, we assume the cluster has been formed, and relay
selection is random. In the future, we focus on how to form a cluster
in PLs efficiently and may propose an appropriate relay selection algorithm
based on car's leaving patterns.

\bibliographystyle{unsrt}
\bibliography{reference/ParkedVehicles/ParkedVehicles}

\begin{IEEEbiography}[{\includegraphics[height=1.25in]{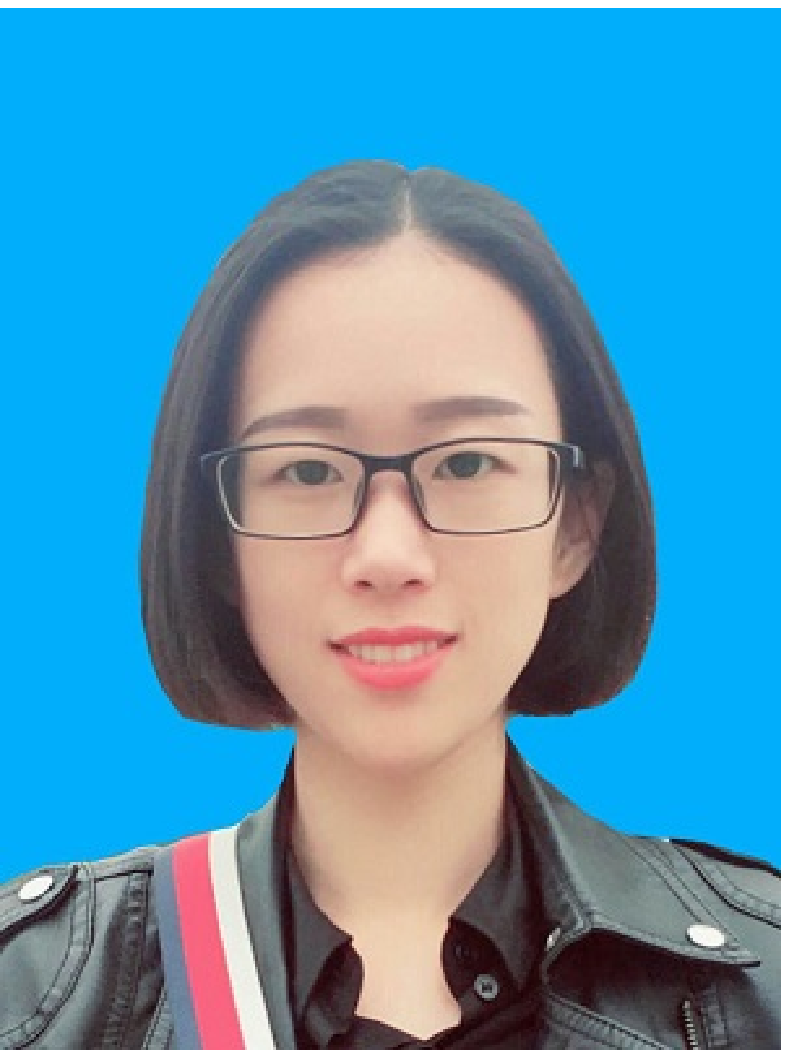}}]{Yingying Sun}
received the Bachelor\textquoteright s degree in electronic and
information engineering from Wuhan University of Technology, Wuhan,
China, in 2016. She is currently working toward the Master\textquoteright s
degree in Huazhong University of Science and Technology (HUST), Wuhan,
China.
Her research interests include cooperative communication and vehicular
network. 
\end{IEEEbiography}

\begin{IEEEbiography}[{\includegraphics[height=1.25in]{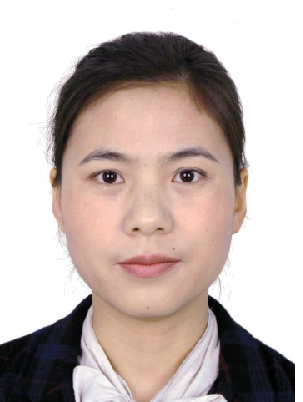}}]{Lijun Wang}
received the B.S. degree in telecommunication
engineering from Xidian University, Xi\textquoteright an, China in
July, 2004 and the M.S. degree in communication and information system
from Huazhong University of Science and Technology (HUST), Wuhan,
China in June, 2008. From Sept., 2016 she has been working toward the Ph.D.
degree with Wuhan University, Wuhan, China.

She is currently an associate professor with the Department of Information
Science and Technology, Wenhua College, Wuhan, China. Her research
interests include wireless communications, and multimedia communications.
\end{IEEEbiography}

\begin{IEEEbiography}[{\includegraphics[height=1.25in]{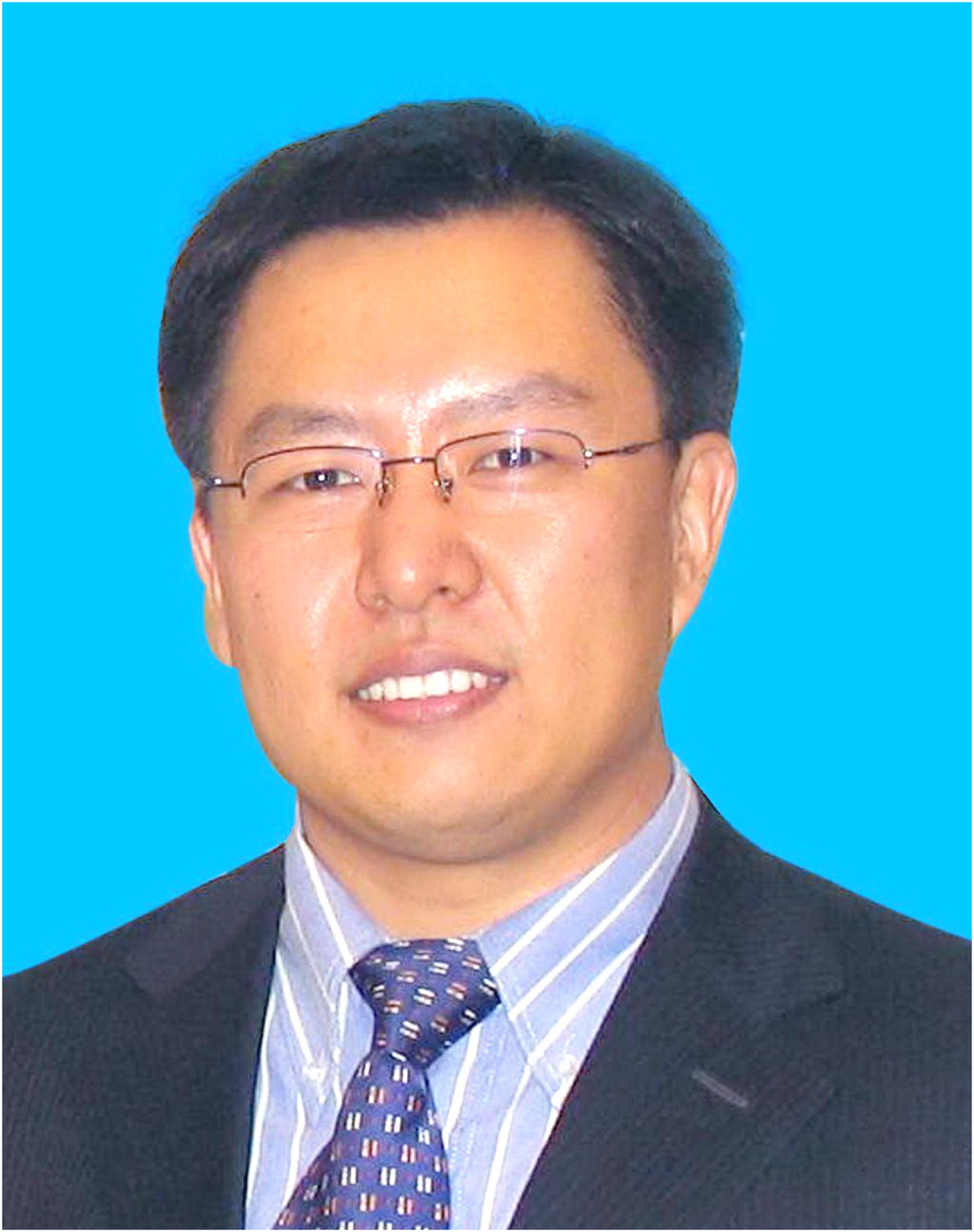}}]{Zhiquan Bai} 
received the M.Eng. degree in communication
and information systems from Shandong University, Jinan, China, in
2003, and the Ph.D. degree in communication engineering with honor
from INHA University in 2007, under the Grant of Korean Government
IT Scholarship, Incheon, Korea.

From 2007 to 2008, he was a postdoctor with UWB Wireless Communications
Research Center, INHA University, Incheon, Korea. Since 2007, he has
been an associate professor with the School of Information Science
and Engineering, Shandong University, China. He has published more
than 70 papers in international conferences and journals. His current
research interests include cooperative communications and MIMO system,
ultra wideband communications, cognitive radio system, and beyond-fourth
generation wireless communications.

Dr. Bai is the associate editor of the \emph{Introduction Journal
of Communication Systems}. He served as a TPC member and a session
chair for some international conferences.
\end{IEEEbiography}

\begin{IEEEbiography}[{\includegraphics[height=1.25in]{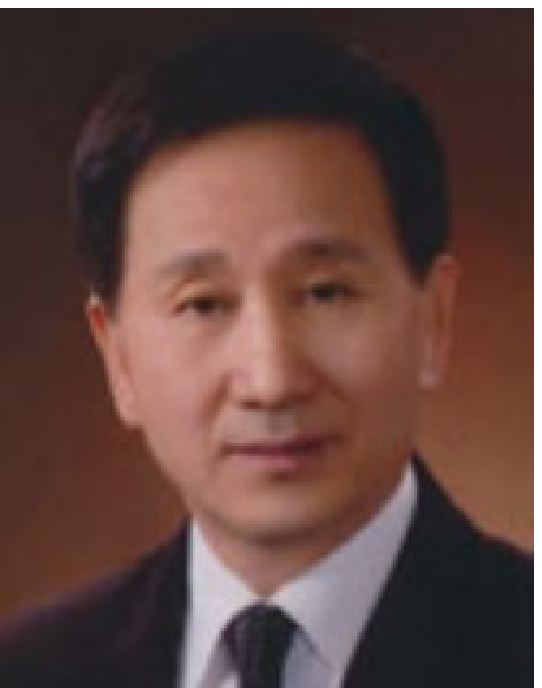}}]{Kyung Sup Kwak} 
received his BS degree from the Inha University, 
Inchon, Korea, in 1977 and his MS degree from the University of Southern
California in 1981 and his PhD degree from the University of California at 
San Diego in 1988, under the Inha University Fellowship and the Korea 
Electric Association Abroad Scholarship Grants, respectively.

Since then, he has been with the School of Information
and Communication Engineering, Inha University,
Incheon, South Korea, as a Professor, and
served as the Dean of the Graduate School of Information
Technology and Telecommunications and the
Director of the UWB Wireless Communications Research
Center, an IT research center, South Korea,
since 2003. In 2006, he was the President of the Korean Institute of Communication
Sciences (KICS), and in 2009, the President of the Korea Institute of
Intelligent Transport Systems. He published more than 200 peer-reviewed journal
papers and served as TPC and Track Chairs/Organizing Chairs for several
IEEE related conferences. His research interests include multiple access communication
systems, mobile and UWB radio systems, future IoT, and wireless
body area networks: nano networks and molecular communications. In 1993,
he received the Engineering College Achievement Award from Inha University,
and a service award from the Institute of Electronics Engineers of Korea, in
1996 and 1999 he received distinguished service awards from the KICS. He
received the LG Paper Award in 1998, and Motorola Paper Award in 2000.
He received official commendations for UWB radio technology research and
development from the Minister of Information and Communication, Prime Minister,
and President of Korea in 2005, 2006, and 2009, respectively. In 2007,
he received Haedong Paper Award and in 2009, Haedong Scientific Award of
research achievement. In 2008, he was elected as an Inha Fellow Professor and
currently as an Inha Hanlim Fellow Professor.

\end{IEEEbiography}

\begin{IEEEbiography}[{\includegraphics[height=1.25in]{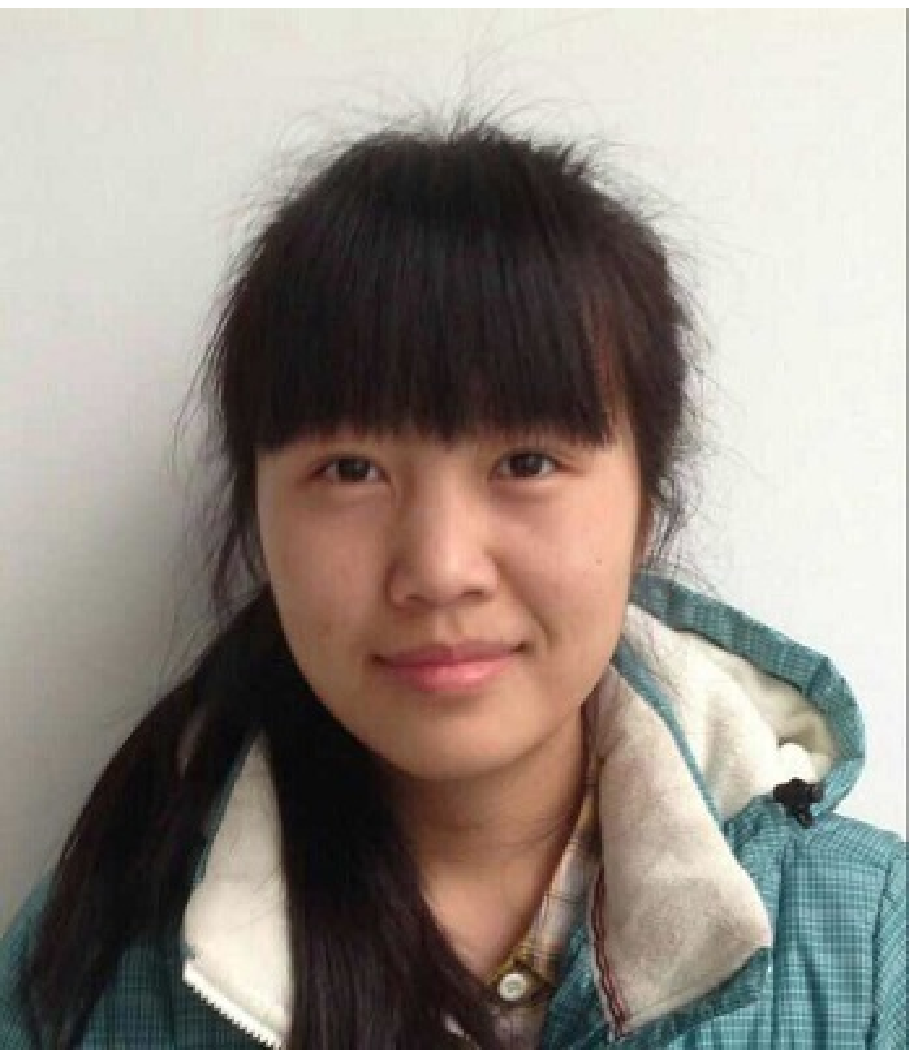}}]{Xuming Yao}
received the Bachelor\textquoteright s degree in electronic and
information engineering from Wuhan University of Technology, Wuhan,
China, in 2016. She is currently working toward the Master's degree in the University of Edinburgh, 
Edinburgh, the United Kingdom.

Her research interests include cooperative communication and signal processing.

\end{IEEEbiography}

\begin{IEEEbiography}[{\includegraphics[height=1.25in]{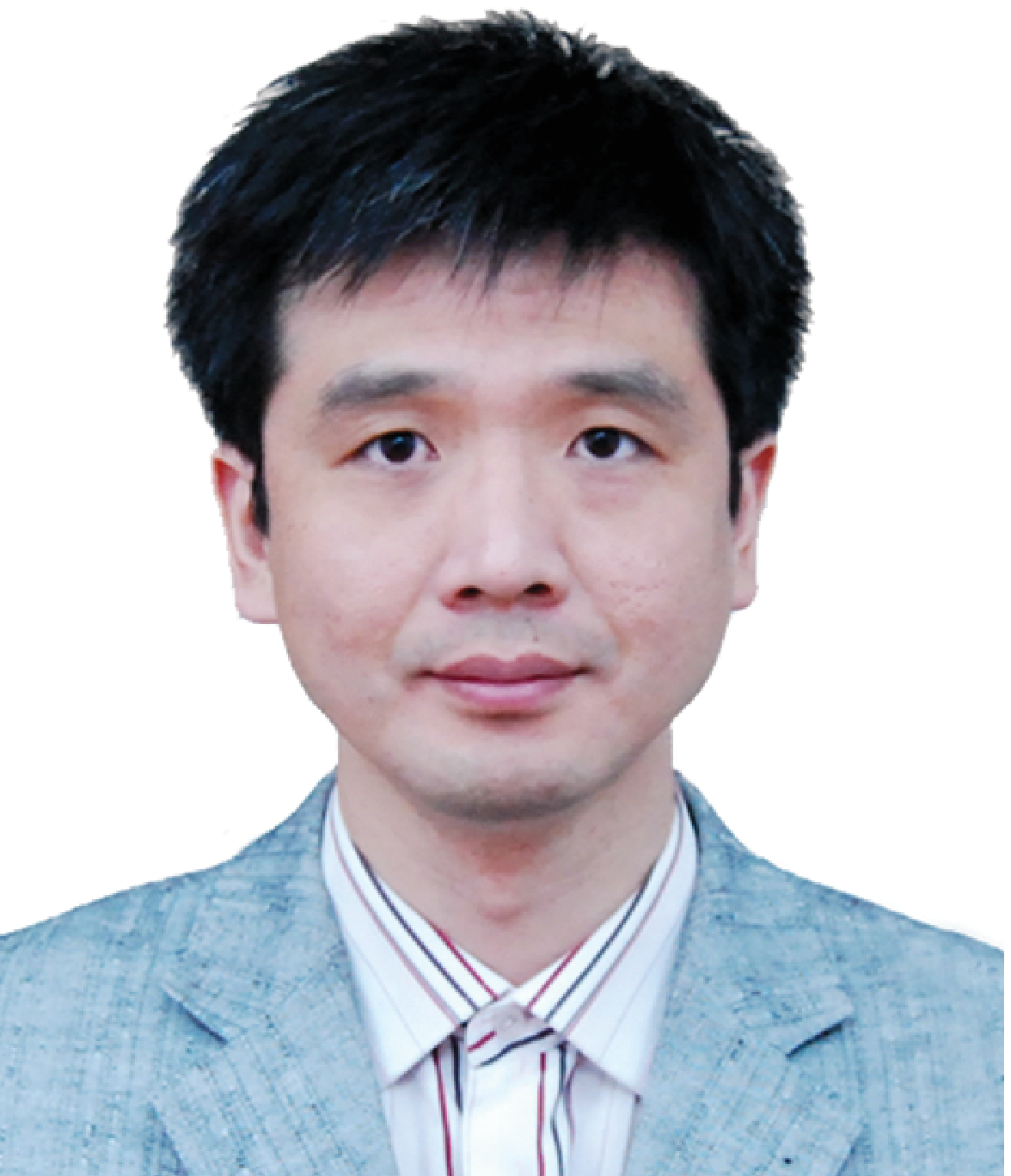}}]{Tao Han} 
received the Ph.D. degree in information and
communication engineering from Huazhong University of Science and
Technology (HUST), Wuhan, China, in 2001.

He is currently an Associate Professor with the School of Electronic
Information and Communications, HUST. From 2010 to 2011, he was a
Visiting Scholar with the University of Florida, Gainesville, FL,
USA, as a Courtesy Associate Professor. He has published more than
50 papers in international conferences and journals. His research
interests include wireless communications, multimedia communications,
and computer networks.

He is currently serving as an Area Editor for the \emph{European Alliance
Innovation Endorsed Transactions on Cognitive Communications}.
\end{IEEEbiography}

\end{document}